# Electronic dynamics in long linear and cyclic polyynes towards the carbyne limit

[1]Soumyadip Bhunia, [2]Yueze Gao, [3,4]Jack Woolley, [1]Ross Milverton, [2]Harry L Anderson, [1]Raj Pandya*

*correspondance: raj.pandya@warwick.ac.uk

[1]Department of Chemistry, University of Warwick, CV4 7AL, Coventry, United Kingdom
[2]Department of Chemistry, University of Oxford, Chemistry Research Laboratory, Oxford OX1 3TA, U.K.
[3]Department of Physics, University of Warwick, CV4 7AL, Coventry, United Kingdom
[4]Warwick Centre for Ultrafast Spectroscopy, University of Warwick, Coventry, UK

**Abstract**
Carbyne—the one-dimensional sp-hybridised allotrope of carbon—has long been predicted to exhibit unique properties, yet its synthesis remains elusive. To probe its behaviour, finite sp-carbon chains such as cumulenes and polyynes have been studied, but work to date has focused almost exclusively on short, linear systems far from the infinite carbyne limit and without considering topology. Here, we investigate long (48-carbon) linear and cyclic polyynes using steady-state and ultrafast, temperature- and polarization-resolved optical and vibrational spectroscopy. We find highly delocalized ground states in both topologies, with Peierls distortions markedly weaker than in short chains. In contrast, excited states undergo rapid self-localisation, with the localisation pathway and subsequent intersystem crossing strongly dependent on chain length and topology. Unlike shorter polyynes, excited-state structural rearrangements are minimal, and comparison with theoretical predictions shows that properties, such as Huang–Rhys factors, have plateaued by 48 carbons. Our results reveal how topology influences the electronic dynamics of long polyynes and refines our understanding of sp-carbon systems approaching the carbyne limit.

**Introduction**
While two- and three-dimensional allotropes of carbon, *e.g.*, graphite and diamond, are well studied, the simplest one-dimensional carbon allotrope, carbyne, is not.[1–4] Carbyne, whose existence remains controversial, has attracted huge amounts of interest for its: (a) predicted outstanding thermal and electronic properties, making it a candidate for molecular wires; (b) stiffness and toughness that are suggested to exceed those of diamond; and (c) role as a transient intermediate in a variety of chemical reactions.[5–8]

Carbyne, could potentially exist in two forms with sp-hybridized carbon atoms: either as semiconducting polyyne (with alternating single and triple bonds) or as metallic cumulene (where all carbon atoms are connected *via* double bonds).[9] As a route to studying the yet to be synthesised, infinitely connected, carbyne, model compounds with finite chain lengths, polyynes and cumulenes, have been a promising avenue.[10,11]

Over the last 50 years more effort has been placed towards exploring polyynes, due to their increased stability as compared to cumulenes.[12,13] Initially this focused on the synthesis of relatively short (<6 alkyne units) linear systems, which could be stabilized against polymerization in the liquid state *via* the addition of bulky capping groups.[14,15] However, more recently the groups of Gladysz, Anderson, Tykwinski and others have demonstrated routes to extending polyyne lengths *via* insulation with rotaxanes, which may or may not interact with the polyyne chain.[16–21] For cumulenes, a similar capping and macrocycle strategy has been employed, although with generally lower chain lengths being achieved (the longest cumulene studied has had 9 repeating double bond units).[22,23]

The above synthetic developments have allowed for the investigation of the physical properties of polyynes and cumulenes. A hallmark of both is bond-length alternation (BLA), where the difference between adjacent carbon bonds, deviates from the expected values ($\Delta r_{BLA\ polyenes}$ ~0.13–0.19 Å; $\Delta r_{BLA\ cumulenes}$ ~0.01–0.10 Å).[24,25] The BLA is an order parameter for Peierls distortions which open a π-band gap in polyyenes/cumulenes and hence is a quantitative bridge between structure (backbone geometry) and electronic properties in polyenes/cumulenes.[26,27] It has, for instance, been shown that $\Delta r_{BLA}$ drops with the number of alkyne units in polyynes, resulting in smaller bandgaps.[28] From such studies it has also been possible to predict the expected properties of polyynes in the infinite ($\lambda_\infty$ = 503 nm, Raman frequency of $v_\infty$ = 1,900 cm$^{-1}$) and carbyne limit (bandgap >1.6 eV).[16,29,30]

Beyond static electronic structure, a few works have started to address electronic *dynamics* in polyynes and cumulenes.[31–33] Optical absorption measurements have established that there are two absorption regimes for polyynes, Abs$_{main}$ is strongly dipole-allowed (S$_0$–S$_n$), while Abs$_{weak}$ arises from a weak S$_0$–S$_{2/3}$ transitions (which become forbidden as chains become longer). Ultrafast spectroscopy measurements on short chains (<10 alkyne units) have indicated that photoexcitation into dipole-allowed S$_n$ states is followed by ~200 fs internal conversion to S$_1$, and slow triplet formation *via* picosecond intersystem crossing.[34] Where in polyynes photoexcitation induces a cumulenic reorganization in the excited state structure, in cumulenes the opposite occurs *i.e.*, polyynic character is induced by photoexcitation.[35] In these latter systems with a ground-state cumulene structure, the excited state dynamics are dominated by a rich-set of interactions with the end-groups that are used to stabilize the cummulene.

Despite the outstanding progress detailed above, much remains unknown about sp-conjugated carbon allotropes and how their electronic structure/dynamics relate to theoretical predictions.[36] Firstly, all the advanced spectroscopic measurements to-date have been performed on relatively short sp-carbon systems e.g., for polyynes <18 alkyne units.[33,37–39] How the underlying electronic dynamics of sp-carbons, be they polyynic or cumulenic, evolve in the infinite carbyne limit remain experimentally unknown. Secondly, it remains unclear how topology influences the properties of sp-conjugated carbons. While discussion has focused on the potential properties of *linear* carbyne, it is possible also to envisage a cyclic structure. Indeed, the equivalent cyclic polyynes (termed cyclocarbons) and cumulenes have been theoretically predicted to host a variety of additional, unusual electronic and structural properties with suggestions that they may play an important role in interstellar environments as intermediates in the formation of prebiotic carbonaceous compounds. Yet stabilizing cyclic polyynes or cumulenes beyond the gas phase or at cryogenic temperatures (and only as small ring structures) has remained challenging. Consequently, most of our understanding of how topology influence the properties of sp-carbons has been limited to on-surface structural studies or *via* computational modelling.[40,41]

Here, we overcome the above challenges and explore the electronic structure and dynamics of sp-carbons towards the carbyne limit (>22 alkyne units), comparing both linear and cyclic geometries. Across both geometries, BLA remains the key parameter controlling the optical gap, with linear chains exhibiting larger gap–BLA coupling due to stronger Peierls distortion and larger excited-state polarizability shifts. Crucially, unlike short polyynes, the BLA in long chains is no longer dictated by localized C≡C stretches, revealing a highly delocalized ground state with substantial cumulene character. Upon photoexcitation, singlet excitons in long polyynes undergo self-trapping with minimal perturbation from side groups. These singlets undergo a subsequent intersystem crossing which occurs more slowly in linear chains compared to cyclic polyynes. Strikingly, the singlet excitons in the two topologies depolarize via distinct mechanisms: linear polyynes depolarize through rapid state mixing and dipole reorientation during internal conversion, whereas cyclic systems show negative initial anisotropy and faster depolarization driven by electronic reorganization within the ring. Interestingly, in contrast to short polyynes, we find the structural reorganisation on photoexcitation in long polyynes to be both very short-lived (sub-5 ps) and small, albeit towards a more cumuluene-like structure. Our results answer long-standing questions about how the properties sp-carbons are influenced by chain length and topology towards the infinite carbyne limit, suggesting where and where not extrapolations can be made from the more readily studied short chain oligoynes.

**Results**

Synthesising long-polyyne chains in both linear and cyclic geometries is challenging. However, several groups have demonstrated solutions to the problem, cumulating in the synthesis in 2025 of a cyclic 24-alkyne polyyne, cyclo[48]carbon, that is stable in solution with a half-life of about 92 hours.[17] In this work we focus on linear[48]carbon and cyclo[48]carbon polyynes, which we henceforth refer to as **C$_{48}$-chains** and **C$_{48}$-rings** (see **Figure 1**). Synthesis follows identically to the previous work of Gao et al.. Linear **C$_{48}$-chains** were stabilised using tris(3,5-di-*tert*-butylphenyl)methyl terminals and threaded phenanthroline macrocycles [16] and **C$_{48}$-rings** were stabilised using threaded macrocycles based on 2,2'-bipyridine.[17]

The absorption spectra of the long linear and cylic polyynes are dominated by the one-photon–allowed π→π* transitions (doubly degenerate in the case of **C$_{48}$-rings**). As shown in **Figure 2a** this appears as a broad UV–vis band bearing a clear Franck–Condon progression. Analysing the spacing of the progression in **C$_{48}$-chains** and **C$_{48}$-rings**, we find vibronic peak spacings of around ~0.2 eV in both (1610 cm$^{-1}$). This is much smaller than in short-polyynes where the spacing is around 0.25 eV (2000 cm$^{-1}$) i.e., close to the C≡C stretching frequency.[18] This deviation suggests in longer polyynes the dominant vibronic mode is no longer localized, but a lower-energy collective mode associated with reduced Peierls distortion (BLA) and increased π-electron delocalization along the chain.

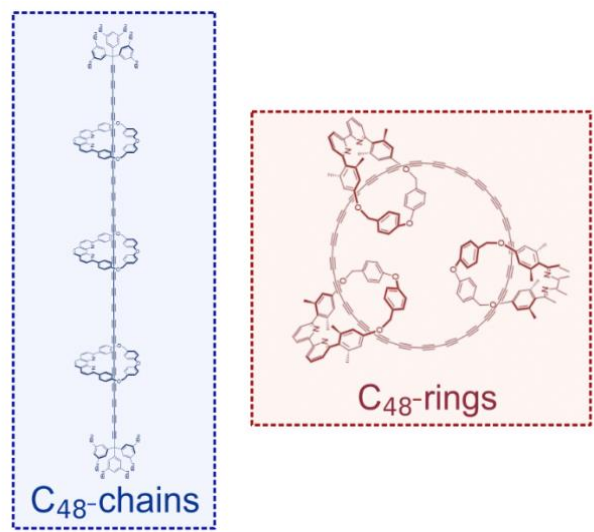

**Figure 1: Chemical structure of C$_{48}$-rings and C$_{48}$-chains studied in this work.**

A second quantity accessible from the UV-Vis spectra in **Figure 2a** is the Huang–Rhys (HR) factor which represents the average number of quanta involved in the vibrational transition and measures the strength of electron-phonon coupling.[42] Given the near constant spacing between vibronic peaks in **Figure 2a**, we can apply the following relation $\frac{I_{0\to 0}}{I_{0\to v}} = \frac{S^v}{v!}$ to determine the Huang-Rhys factor $S$ (this relation states that $v^{th}$ power of the HR factor $S$ is proportional to the ratio of the intensity of the $v^{th}$ vibronic transition compared to the fundamental transition).[43] We find $S$ values of 0.95 and 1.21, respectively for **C$_{48}$-chains** and **C$_{48}$-rings**. For **C$_{48}$-chains**, the value is especially close to that found in shorter chains (e.g., 22 carbon polyynes where $S$ = 0.95), and much smaller than the theoretical value calculated for infinite chains (1.82).[43] Firstly, this suggests that $S$ may in-fact experimentally plateau before the theoretically predicted value of 1.82 due to e.g., the diminishing influence of Peierls distortions, beyond a critical polyyne chain length. Secondly, the fact that $S$<1 for **C$_{48}$-chains** but $S$>1 for **C$_{48}$-rings** indicates that while in both excitation will result in limited structural change (due to the low values of $S$), in the latter any changes will be larger, which, in turn, suggests that the cyclic topology exhibits more carbyne-like behaviour, consistent with its HR factor lying closer to the infinite-chain limit.

To explore further how BLA varies between the **C$_{48}$-chains** and **C$_{48}$-rings**, we performed variable-temperature absorption measurements, in combination with temperature-dependent Raman around the C≡C stretch. Although the BLA is not dominated by only this mode, it remains a sensitive reporter of changes in electronic delocalization. Turning to the Raman spectra first, as shown in **Figure 2b**, on cooling both **C$_{48}$-chains** and **C$_{48}$-rings** from 300 K to 190 K there is a shift of the main C≡C stretch from ~1890 cm$^{-1}$ to ~1910 cm$^{-1}$ for **C$_{48}$-rings**, whereas for **C$_{48}$-chains** the shift is smaller from ~1890 cm$^{-1}$ to ~1900 cm$^{-1}$ (even after solvent correction).

We can couple these observations to the temperature-evolution of the band-edge, computing a simultaneous, two-predictor fit of the optical band edge to the C≡C Raman wavenumber and the solvent function (to disentangle intrinsic backbone effects from solvent-induced shifts),

$$E_{edge}(T) = a + b[\nu_{C\equiv C}(T) - \nu'] + c[\Phi(T) - \Phi(T_{ref})]$$ (**Equation 1**).

Here, $E_{edge}(T)$ is the optical band, $\nu_{C\equiv C}(T)$ the triple bond stretch frequency at a given temperature, $a$ the point when $\nu_{C\equiv C}(T) = \nu'$ and $\Phi(T) = \Phi(T_{ref})$, $b$ the gap-BLA sensitivity, $\nu'$ the Raman frequency of the C≡C stretch at the lowest measured temperature and $\Phi(T)$ the McRae/Bakhshiev polarity–polarizability function of the solvent that allows for correction of the UV-Vis (see **Supplementary Note 1**). **Equation 1** is then evaluated by performing a simultaneous least-squares regression over all temperatures, yielding the gap–BLA sensitivity $b$ and solvent sensitivity $c$ as global fitting parameters. To visualise the intrinsic coupling between BLA and the electronic gap, **Figure 2c** shows the solvent-corrected band edge, $E_{edge}(T) - c_{fit}[\Phi(T) - \Phi(T_{ref})]$, plotted as a function of the referenced C≡C stretching frequency, $\nu_{C\equiv C}(T) - \nu'$. The resulting linear relationships confirm that, despite the softened BLA in long polyynes, bond-length alternation remains the dominant order parameter controlling the band-edge energy in both linear and cyclic topologies. Notably, both the extracted gap–BLA sensitivity $b_{fit}$ (2.8 × $10^{-3}$ eV cm$^{-1}$ for chains and 1.0 × $10^{-3}$ eV cm$^{-1}$ for rings) and solvent sensitivity $c_{fit}$ (0.5 eV for chains and 0.17 eV for rings) are larger for **C$_{48}$-chains** than for **C$_{48}$-rings**, indicating a stronger coupling between local backbone distortions and the π–π* gap, as well as a larger excitation-induced change in electronic polarizability ($\Delta\alpha$), in the linear system.[44] This reduced sensitivity in **C$_{48}$-rings** can be rationalised by the presence of two orthogonal, degenerate π-conjugation pathways around the ring, which delocalise the electronic response and thereby dilute the impact of local structural or dielectric perturbations on the global band-edge energy.

The $b_{fit}$ values for **C$_{48}$-chains** and **C$_{48}$-rings** can be converted to more general values for the gap-BLA sensitivity ($\frac{dE_{edge}}{d(\Delta r)}$), by noting that $\frac{dE_{edge}}{d(\Delta r)} = b_{fit} \times k_r$, with $k_r = \frac{d\nu_{C\equiv C}}{d(\Delta r)}$ calculated using literature values.[24] Overall, we find $\frac{dE_{edge}}{d(\Delta r)}$ values of 11.9 eV Å$^{-1}$ (rings) and 31.54 eV Å$^{-1}$ (chains). We note that these values are smaller than those reported for long polyyne chains encapsulated in carbon nanotubes and theoretical diffusion Monte Carlo calculations.[45,46] However, our values potentially reflect a more intrinsic gap–BLA coupling strength where environmental factors such as dielectric screening by a nanotube encapsulant do not influence the obtained values.

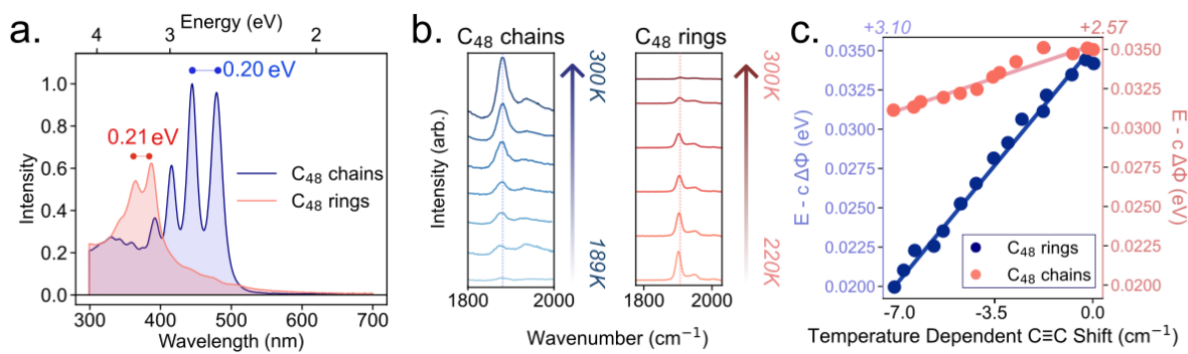

**Figure 2: Ground-state electronic structure and bond-length alternation in long linear and cyclic polyynes a.** Room-temperature absorption spectra of **C$_{48}$-rings** and **C$_{48}$-chains**, all recorded in dichloromethane solvent. The vibrational progression of 0.2 eV is marked. **b.** Temperature-dependent Raman spectra of **C$_{48}$-rings** and **C$_{48}$-chains**, both show a small 7 cm$^{-1}$ shift on cooling from room temperature to 200 K. The dotted lines indicate the Raman peak maxima of **C$_{48}$-rings** and **C$_{48}$-chains** at room temperature. **c.** Solvent-corrected optical band-edge energy plotted as a function of the referenced C≡C stretching frequency, $\nu_{C\equiv C}(T) - \nu'$, for **C$_{48}$-chains** (blue) and **C$_{48}$-rings** (red), with each data point corresponding to a different temperature. The band-edge energies are corrected for solvent polarity–polarizability effects using the fitted solvent sensitivity $c_{fit}$ (**Equation 1**). Solid lines show the best-fit linear regressions obtained from the simultaneous least-squares analysis, with the slopes yielding the gap–bond-length-alternation (BLA) sensitivities $b_{fit}$.

Having captured the ground-state properties of long linear and cyclic polyynes, we turn to their excited state dynamics using ultrafast transient absorption spectroscopy. **Figure 3a,b** displays the ultrafast transient absorption response of **C$_{48}$-chains** and **C$_{48}$-rings**, pumped at 400 nm (sealed solutions unless otherwise stated; see methods). Here, ΔmOD is the pump-induced change in optical density, defined as $\Delta mOD = -1000\log_{10}[(T + \Delta T)/T]$, where $\Delta T$ is the change in the transmission of the sample induced by the photoexciting pump pulse and $T$ is the transmission in the absence of the pump. ΔmOD is plotted as a function of probe wavelength and pump–probe time delay. In both **C$_{48}$-chains** and **C$_{48}$-rings** the spectra display a strong ground state bleach (GSB; negative ΔmOD signal) between 320 and 400 nm (500 nm in **C$_{48}$-chains**), with a broad photoinduced absorption band in the near-infrared (positive ΔmOD signal). No stimulated emission, which would appear a positive ΔmOD signal red-shfited from the GSB, is observed for either molecule. This verifies the theory that regardless of the polyyne chain length or topology the S$_1$ state is dark, with fast non-radiative relaxation. In the GSB region for **C$_{48}$-chains** and **C$_{48}$-rings** a small blue-shift is observed in the first 300 fs following photoexcitation (~25 meV for **C$_{48}$-chains** and ~15 meV **C$_{48}$-rings**; see **Supplementary Note 2** where we zoom into the spectra at short pump-probe delays). Such behaviour would be in-line with self-trapping along the BLA coordinate—i.e., a rapid increase in bond-length alternation that opens the gap (blue-shifting the GSB) and funnels population into a dark S$_1$ manifold. The larger shift for **C$_{48}$-chains** as opposed to rings **C$_{48}$-rings** is consistent with the larger BLA sensitivity as demonstrated from the temperature dependence of the band-edge and C≡C bond frequency.

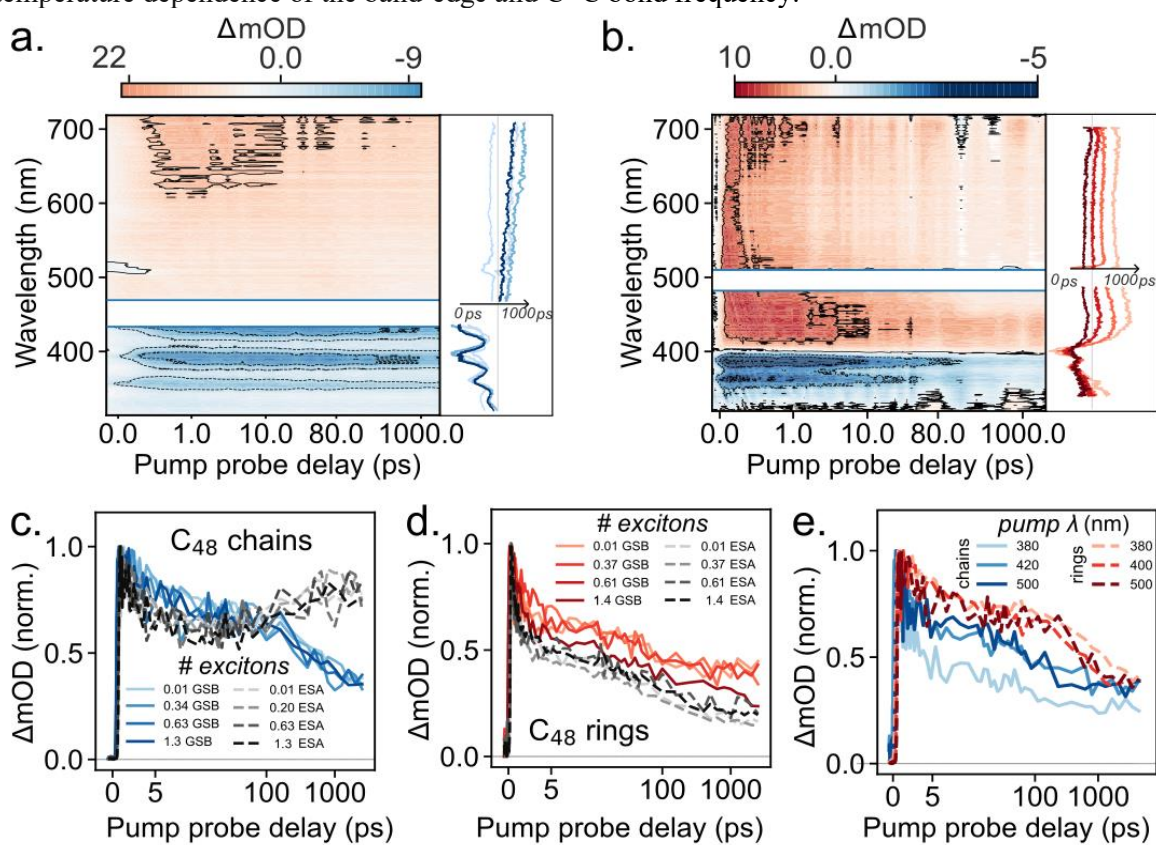

**Figure 3: Ultrafast electronic dynamics of long cyclic and linear polyynes: a-b.** Ultrafast transient absorption spectra of **C$_{48}$-chains** (a) and **C$_{48}$-rings** (b) following 480 nm and 500 nm excitation. The pump region has been blocked out in white in both figures. **c-d.** Ground-state bleach (GSB) and excited state absorption (ESA) dynamics of **C$_{48}$-chains** and **C$_{48}$-rings** measured as a function of pump fluence, corresponding to excitation densities (inset; **C$_{48}$-chains** 0.01 to 1.3 and **C$_{48}$-rings** 0.01 to 1.4). **e.** GSB dynamics measured as a function of pump wavelength (380, 420 and 500 nm) showing minimal dependence on excitation energy for both chain and ring topologies. All experiments were performed in dichloromethane solvent at room temperature.

To explore the hypothesis of self-trapping further, we perform transient absorption measurements as a function of pump fluence in **Figure 3c,d**. Varying the number of excitations per molecule in **C$_{48}$-chains** and **C$_{48}$-rings** up to 1.4 excitons per molecule, we see little change in the GSB dynamics.[47] This behaviour is consistent with rapid localization into small polarons on the BLA (Peierls) coordinate that quench mobility, such that even at high excitation densities excitons do not encounter one another. Only at 1.4 excitations per molecule in the **C$_{48}$-rings** specifically, is a small, but consistent change GSB lifetime observed as shown by the marron kinetic in **Figure 3d**. This suggests in the rings there may be a small amount of annihilation and excited state delocalisation.[48,49] We note a similar insensitivity to pump wavelength is also observed, with pumping closer to the band edge resulting in a slight increase in the GSB decay in **C$_{48}$-rings**, but generally kinetics that do not depend on the pump wavelength in **C$_{48}$-rings** or **C$_{48}$-chains** (see **Figure 3e**). The strong exciton self-trapping we observe contrasts with shorter polyyne chains and other nanoring/nanostring type molecules based on e.g., porphyrin or diacetylene units, where significant exciton delocalisation has been observed. These systems have stiffer backbones, weaker electron–phonon coupling, and stronger collective excitonic coupling which likely promote singlet exciton diffusion.[50,51]

The absence of pump-wavelength dependence in our transient absorption kinetics also rules out energy transfer from the **C$_{48}$-rings** or **C$_{48}$-chains** to their stabilising macrocycles. The macrocycle units absorb in the 310–380 nm region, overlapping with the polyyne absorption, such that UV excitation in this range can directly excite both the macrocycle and the polyyne.[33] However, the transient spectra and kinetics remain essentially identical across this wavelength range and are consistent with those obtained at redder pump wavelengths where excitation is dominated by the polyyne alone. If energy transfer to the macrocycle or any macrocycle-mediated decay pathway were operative, additional wavelength-dependent kinetic components or accelerated relaxation would be expected. The lack of such effects indicates that no new decay channels are introduced by macrocycle excitation and that the macrocycles are effectively inert with respect to the observed excited-state dynamics.

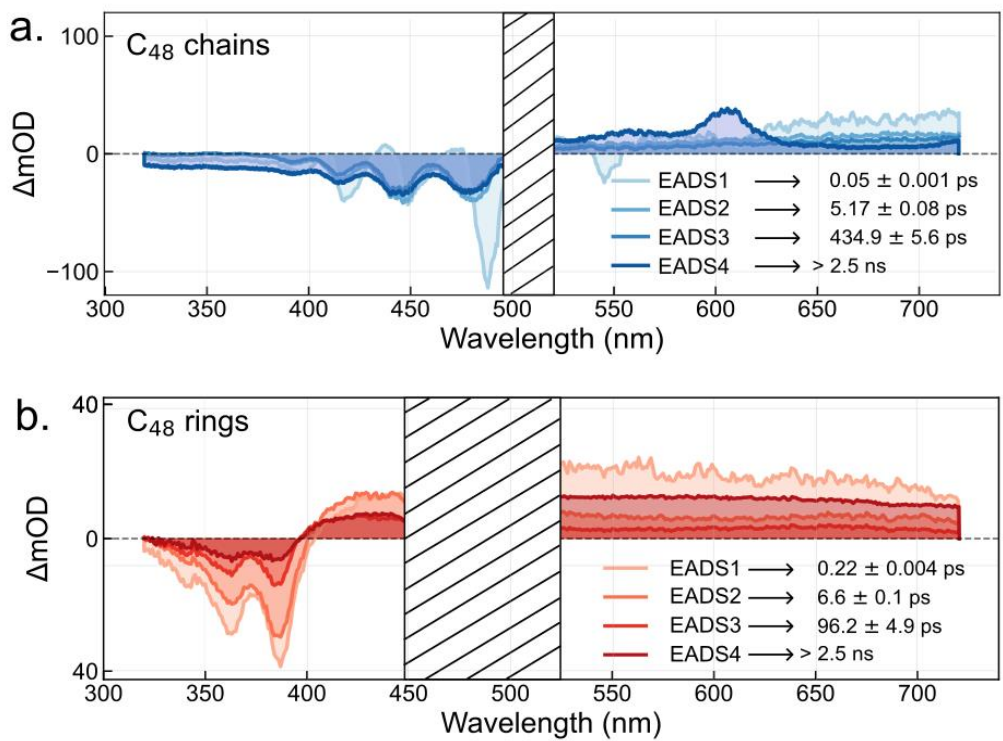

**Figure 4: Evolution-associated decay spectra of C$_{48}$-chains and C$_{48}$-rings.** Decay-associated spectra obtained from global analysis of the transient absorption data for **C$_{48}$-chains** (a) and **C$_{48}$-rings** (b), illustrating the successive relaxation of hot singlet states, formation of the relaxed S$_1$ state, and population of the triplet manifold via intersystem crossing. Each EADS corresponds to the spectral amplitude associated with a distinct kinetic time constant indicated in the legend. All experiments were performed in dichloromethane solvent at room temperature.

We next turn to assigning the excited-state electronic manifold of the long linear and cyclic polyynes. To this end, we apply global analysis to the transient absorption spectra in **Figure 4a,b** to obtain evolution-associated decay spectra (EADS).[52] EADS represent the spectral amplitudes associated with each kinetic time constant in the fitted model and describe how the transient spectral features evolve over time. While EADS do not correspond to pure species spectra, their spectral signatures and lifetimes provide insight into the characteristic timescales and pathways of the underlying excited-state dynamics.

As shown in **Figure 4a,b** (and **Supplementary Note 3)**, four components are required to reproduce the transient absorption dynamics, yielding qualitatively similar spectral components for **C$_{48}$-rings** and **C$_{48}$-chains**. The fastest component (EADS1) is spectrally broad and exhibits a predominantly positive ΔmOD feature centred at 600 nm and 650 nm for **C$_{48}$-rings** and **C$_{48}$-chains**, respectively. Given that photoexcitation in polyynes initially accesses high-lying singlet states, and that EADS1 is associated with a very short lifetime (0.1 ps for chains and 0.3 ps for rings), we attribute this component primarily to excited-state absorption originating from the initially prepared singlet population (hot S$_1$ or higher-lying singlet states) prior to internal conversion. The second component (EADS2) has a lifetime of 3.3 ps for the rings and 5.1 ps for the chains and is characterised by: (i) a negative feature near the steady-state absorption maximum and (ii) a positive band that is blue-shifted relative to that observed in EADS1. In light of the rapid exciton localisation previously identified in both systems—occurring on a timescale comparable to the decay associated with EADS1—we assign EADS2 to the relaxed S$_1$ state. The associated spectral evolution reflects the structural and electronic relaxation accompanying exciton localisation.

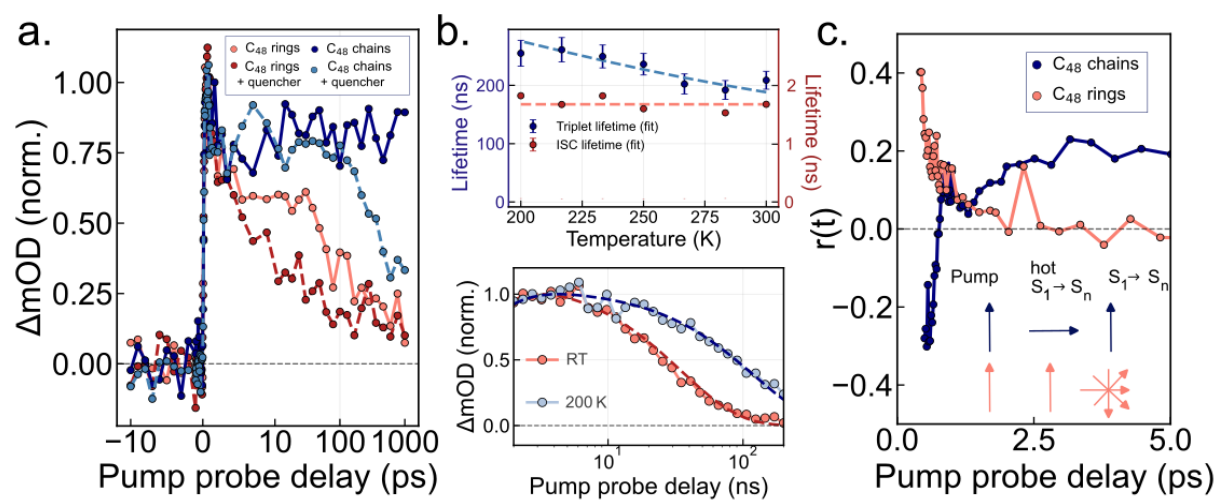

Figure 5: **Topology-dependent intersystem crossing and ultrafast depolarisation dynamics in long polyynes. a.** Comparison of the ESA kinetics of **C$_{48}$-rings** and **C$_{48}$-chains** with and without a quencher. **b.** Nanosecond time resolution (~0.7 ns) transient absorption spectroscopy experiments at room and low-temperature on **C$_{48}$-chains** (bottom panel). Fitting the kinetics monitored at the ESA band (top panel) with a biexponential function allows extraction of two time-constants which we assign to the ISC and triplet lifetime. Plotting these lifetimes as a function of temperature shows that the ISC lifetime (right y-axis) is independent of temperature, whereas the triplet lifetime (left y-axis) is strongly dependent on temperature (the straight line is a line of best fit through the lifetimes as a function of temperature). **c.** Transient absorption anisotropy dynamics measured on the excited-state absorption (ESA) for **C$_{48}$-rings** and **C$_{48}$-chains**. In **C$_{48}$-rings**, the ESA anisotropy initially reaches the theoretical maximum ($r \approx 0.4$) and decays to zero within ~2.5 ps, indicating rapid depolarisation driven by vibronic and electronic relaxation that mixes the nearly degenerate, orthogonal π states of the cyclic backbone. In contrast, **C$_{48}$-chains** exhibit an initial negative anisotropy ($r \approx -0.2$), followed by recovery to a positive steady-state value ($r \approx 0.2$) over ~5 ps, reflecting exciton cooling and population funnelling into a backbone-aligned S$_1$ manifold (cartoon insert).

EADS3 exhibits markedly different lifetimes for rings and chains. In **C$_{48}$-rings**, EADS3 has a lifetime of 93 ps, whereas in **C$_{48}$-chains** it is significantly longer at 431 ps. Previous studies on shorter polyynes have shown that, following photoexcitation, the S₁ state can undergo intersystem crossing (ISC) to a triplet state.[39] The spectrally flat character of EADS3, together with the rising kinetics in the 540-680 nm ESA region (especially evident for **C₄₈-chains**) suggests a similar phenomenon may be occurring here.

To clarify whether triplets are forming via ISC in these long polyynes, and why the formation/decay times differ so markedly between chains and rings, we performed several additional experiments. Firstly, we repeated the transient absorption measurements shown in **Figure 3** in the presence of β-carotene (see **Methods**), a triplet quencher. As shown in **Figure 5a**, addition of the quencher leads to a substantial reduction in the ESA lifetime for both **C₄₈-rings** and **C₄₈-chains** within our 0.01–1 ns detection window. For the chains, this quenching is particularly pronounced, directly indicating that a significant fraction of the long-lived signal arises from a triplet population. Consistent with this, no compensating growth is observed in the 540–680 nm region at times >0.5 ns, which would be expected if the ESA were instead fed by slow singlet relaxation or delayed formation of a distinct long-lived species. We note that experiments where solutions of **C₄₈-rings** are purposefully oxygenated to quench triplets, produce similar results to those shown in **Figure 5a** with addition of β-carotene, and have distinct kinetics from **C₄₈-ring** solutions that are partially decomposed (see **Supplementary Note 3**).

For the more photostable **C₄₈-chains**, we additionally can carry out temperature-dependent transient absorption measurements. As shown in the bottom panel of **Figure 5b**, cooling results in a clear increase in the lifetime of the long-lived ESA component, behaviour characteristic of triplet states formed via intersystem crossing that exhibit extended lifetimes at low temperature. Indeed, as the long-lived ESA dynamics are independent of excitation fluence over the range investigated, we rule out intermolecular singlet fission, which would be expected to exhibit a fluence-dependent triplet yield. Intramolecular singlet fission is also unlikely, because despite the extended conjugation, we observe no evidence for ultrafast formation of a correlated triplet-pair state, with spectra more consistent with a one-to-one singlet–triplet conversion. The temperature-dependence further excludes simple exciton trapping or backbone conformational relaxation as the origin of the nanosecond signal. Fitting the nanosecond kinetics with a bi-exponential decay we can extract two lifetimes: a short-lived component, which closely matches the ESA rise in **Figure 3** that we assign to ISC (weak temperature dependence), and one that is longer-lived and strongly temperature dependent which we assign (based on the above discussion) to the triplet decay. We find the triplet population decays on timescales of several hundred nanoseconds, significantly longer (double) than those reported for short polyynes such as hexayne, indicating that the triplet lifetime increases with chain length.[33] However, the relatively modest increase suggests that this scaling is not exponential, consistent with reduced electron–phonon coupling that offsets any energy-gap-law-driven acceleration of nonradiative decay as might be expected when the conjugation length increases. Together, these observations demonstrate that **C₄₈-chains** form triplets predominantly via intersystem crossing, with the long-lived ESA arising from the T₁ → Tₙ manifold.

The weak temperature dependence of the ISC is consistent with a spin–vibronically mediated process in which triplet formation is barrierless or near barrierless. Such spin–vibronic coupling is expected to be stronger in cyclic structures due to the curvature and reduced symmetry of the backbone. In linear **C₄₈-chains** (and polyynes more generally), the π and σ orbitals are largely decoupled, which suppresses effective spin–orbit interactions and renders ISC relatively inefficient. Cyclization introduces bending and strain that mix π and σ character and break inversion symmetry, thereby enhancing the effective spin–orbit coupling between the singlet and triplet manifolds. We propose that this increase in spin–vibronic coupling promotes the faster S₁→Tₙ intersystem crossing observed in the rings. In addition, the constrained geometry of the **C₄₈-rings** may modify the triplet potential-energy surface or open additional non-radiative decay pathways, accounting for the shorter triplet lifetime relative to the linear analogue.

Having established the overall fate of excitations in ring and chain polyynes, we turn to examine other mechanistic details of the excitation dynamics. Firstly, we monitor the transient absorption anisotropy along the GSB and ESA of **C₄₈-rings** and **C₄₈-chains**. On ultrafast timescales (sub-5 ps) the anisotropy, calculated as $r(t) = \frac{\frac{\Delta T}{T}_{\parallel} - \frac{\Delta T}{T}_{\perp}}{\frac{\Delta T}{T}_{\parallel} + 2\frac{\Delta T}{T}_{\perp}}$ reflects how the transition dipole moment correlation evolves following excitation (∥ indicates pump and probe polarisations are parallel and ⊥ indicates the two are orthogonal).[53] At $t = 0$, $r$ typically reaches its theoretical maximum of 0.4 for perfectly aligned pump and probe dipoles, whereas values approaching 0 indicate complete loss of orientational correlation. Negative values (down to -0.2) arise when the ESA transition dipole is nearly perpendicular to that of the GSB, reflecting significant exciton reorganisation or coupling between distinct electronic states.[54,55]

For both **$C_{48}$-rings** and **$C_{48}$-chains**, the GSB shows $r(t) = 0$ across the entire measurement window, indicating that any initial anisotropy is erased within the instrument response due to ultrafast internal conversion from higher-lying states to $S_1$, for which the $S_0 \rightarrow S_1$ and $S_n \rightarrow S_1$ transition dipoles are not colinear (see **Supplementary Note 4**). The ESA anisotropy shown in **Figure 5c**, however, reveals striking topology-dependent differences. In the **$C_{48}$-rings**, $r(t)$ initially reaches 0.4 before collapsing to zero within 2.5 ps—the same timescale as the decay of the hot ESA—indicating that depolarisation accompanies cooling from the hot $S_1$ to the cold $S_1$ manifold. We attribute this behaviour to vibronic and electronic relaxation that mixes the two nearly degenerate, orthogonal π states inherent to the cyclic polyyne, producing excited-state superpositions with rapidly varying dipole orientations and thereby removing the initial orientational memory without the need for exciton migration.

In contrast, the linear **$C_{48}$-chains** display an initial negative anisotropy ($r(0) = -0.2$), consistent with the hot $S_1 \rightarrow S_n$ transition dipole being nearly perpendicular to the pumped excitation.[56] As relaxation proceeds, the anisotropy rises over ~5 ps to a constant value of $r(t) \approx 0.2$. This recovery reflects population funneling into the backbone-aligned cold $S_1$ manifold characteristic of a one-dimensional polyyne, thereby restoring positive orientational correlation. Together, these opposing trends show that in cyclic polyynes, electronic degeneracy and symmetry-driven state mixing induce rapid depolarisation, whereas in linear polyynes, exciton cooling reinstates dipole alignment along a single molecular axis.

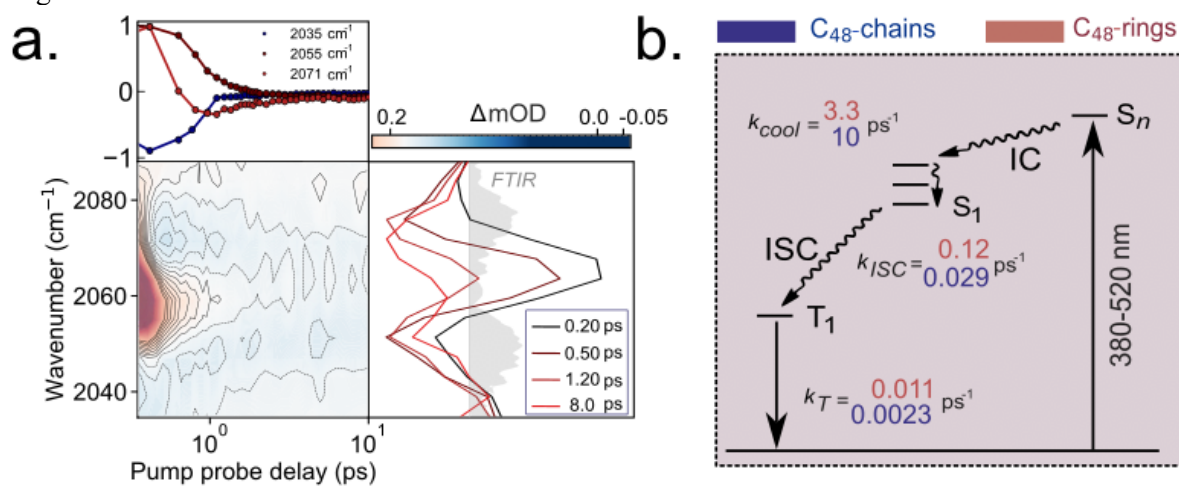

**Figure 6: Ultrafast vibrational dynamics of the polyyne backbone probed by transient infrared spectroscopy and Jablonski diagram summarising rates of excited energy transfer processes. a.** Pump–probe transient IR spectra of **$C_{48}$-chains** following 400 nm photoexcitation, with the probe tuned to the C≡C stretching region (2000–2100 cm$^{-1}$). At early delay times (<3 ps), the spectra exhibit an asymmetric, derivative-like line shape which arises from the superposition of a broad ground-state bleach of the C≡C stretching manifold and a narrower, slightly red-shifted excited-state absorption (ESA) on the $S_1$ surface. Over the subsequent 2–3 ps, the dispersive feature evolves into two negatives ΔmOD bands that sharpen and coincide with the steady-state FTIR absorption peaks (grey shading), indicating vibrational cooling and relaxation from the hot $S_1$ state into the cold $S_1$ minimum. **b.** A Jablonski diagram for energy relaxation in **$C_{48}$-chains** (blue) and **$C_{48}$-rings** (red) can be derived from our transient optical spectroscopy measurements. Using the measured state lifetimes and kinetic modelling, rate constants ($k_{ISC}$ = intersystem cooling rate; $k_{cool}$ = rate for cooling of the hot ground state and $k_T$ = rate for triplet decay) for energy transfer between the excited states can be derived (see **Supplementary Note 5;** red is for rings and blue is for chains).

While the temperature-resolved absorption and Raman measurements in **Figure 2c** establish that both linear and cyclic long polyynes exhibit a softened bond-length alternation (BLA) in the ground state relative to their shorter counterparts, the extent to which this structural softness persists in the excited state requires further investigation. To address this, we perform pump–probe infrared spectroscopy with the probe centred between 2000 and 2100 cm$^{-1}$, encompassing the two IR-active C≡C stretching modes at 2037 and 2082 cm$^{-1}$. Owing to the long acquisition times required for these measurements, the experiment could only be carried out on the more photostable **C$_{48}$-chains**.

Following 400 nm excitation (**Figure 6a**), the transient IR response exhibits an unusual asymmetric, derivative-like line shape that appears within the instrument response (~100 fs) and persists for approximately 3 ps before evolving into two broad negative features that remain throughout the 10 ps measurement window. The early-time dispersive signal can be understood as the superposition of (i) a broad ground-state bleach associated with the two C≡C stretching modes and (ii) a narrower, slightly red-shifted excited-state absorption (ESA) arising from the same vibrational coordinates on the S$_1$ potential energy surface. Because the local bonding framework remains largely intact upon excitation, the S$_1$(v = 0 → 1) transitions occur at frequencies very close to their S$_0$ counterparts, differing only by a small red shift that reflects a transient softening of the backbone as the bond-length alternation is reduced. The close spectral proximity of these contributions leads to partial cancellation near the centre of each bleach band and enhanced negative wings on either side, producing the characteristic negative–positive–negative line shape observed in ΔmOD.

This slight red shift persists on a timescale comparable to that of the hot S$_1$ ESA observed in **Figure 3** and indicates that the initially excited state adopts a geometry that is marginally more cumulene-like than the ground state. This behaviour is consistent with partial equalisation of single and triple bonds, although the magnitude of the structural distortion is much smaller than that observed upon photoexcitation of shorter polyynes or cumulenes.

Over the subsequent 2–3 ps, the dispersive component decays and the spectrum becomes dominated by the two bleach features, indicating vibrational cooling and relaxation from the hot S$_1$ state into the cold S$_1$ minimum. The convergence of the transient spectra with the steady-state FTIR absorption bands (grey shading in **Figure 6a**) confirms that any cumulene-like distortion is short-lived and confined to the initially excited state. In contrast to shorter polyynes and cumulenes—where photoexcitation induces substantial and persistent backbone reorganisation—the relaxed excited state of long polyynes retains predominantly polyyne character, consistent with a weakened but still operative Peierls distortion.

**Conclusion**

In summary, using a combination of steady-state and time-resolved optical spectroscopy, we have provided a detailed picture the electronic properties of long polyynes and how they are influenced by topology (rings versus chains; see **Figure 6b** for a Jablonksi diagram and rate constants for energy transfer between the excited states). Regardless of topology, long polyynes exhibit a softened bond-length alternation relative to their shorter counterparts, reflecting increased π-electron delocalisation and a ground-state geometry that is partially cumulene-like. Although this Peierls softening is more pronounced in linear chains as demonstrated by temperature-dependent absorption and Raman measurements. In contrast to their delocalised ground states, ultrafast optical spectroscopy demonstrates the excited-state dynamics of both long polyyne rings and chains are dominated by rapid self-localisation, which is again greater in linear systems. However, ultrafast anisotropy measurements reveal the mechanism of this trapping differs depending on topology. Whereas in linear chains, following photoexcitation, population is funnelled into a backbone-aligned S$_1$ manifold, recovering positive orientational correlation with the excitation population, the large polyyne rings undergoes rapid depolarisation probably through symmetry-driven mixing of two orthogonal π manifolds. This topological distinction also manifests in their intersystem crossing efficiencies: long polyyne chains readily form triplets, whereas large polyyne rings show markedly faster ISC and shorter triplet lifetimes, consistent with enhanced spin–vibronic coupling in the curved, symmetry-broken cyclic backbone. Our measurements (ultrafast vibrational spectroscopy), free of stabilising side-group effects, also reveal that photoexcitation induces only a very small, transient reduction in the BLA—a slight softening of the C≡C vibration in the hot S$_1$ state—before rapid cooling restores the polyyne-like character of the relaxed excited (S$_1$) state. This contrasts sharply with the pronounced cumulene-like distortions observed in short polyynes and cumulenes, demonstrating that long polyynes occupy a distinct structural regime where strong electron–phonon coupling drives ultrafast localisation but does not fully suppress Peierls-type ordering. Our results take us a step closer to understanding the properties of sp-carbons at the infinite limit. They highlight that at even 48 carbons, in chains and rings, finite size effects remain, albeit weakened (e.g., near-complete retention of polyyne character in the excited state) and even potentially saturated e.g., Huang-Rhys factors. At the same time, they demonstrate that topology shapes the electronic dynamics through spin–vibronic coupling. Future studies should be focussed to studying sp-carbons (polyynes and cumulenes) at greater lengths if suitable stabilisation can be achieved and using strain or other topologies to understand whether carbyne ultimately might behaves as a Peierls-distorted polyyne or a cumulene-like semimetal. More broadly, our findings establish chain length and topology as practical control parameters for tuning intersystem crossing and excited-state electronic dynamics in sp-carbon systems, with direct implications for the design of carbon allotropes, molecular wires, and low-spin–orbit organic materials where spin and charge transport must be precisely engineered.

## Acknowledgements

S.B. acknowledges financial support from the Royal Society Newton Fund (NIF\R1\242215). H.L.A. and Y.G. acknowledge support from European Research Council (grant 885606, ARO-MAT) and the European Commission Horizon 2020 (grant 101019310 CycloCarbonCatenane. The authors gratefully acknowledge the use of the Warwick Centre for Ultrafast Spectroscopy Research Technology Platform (RTP) and the Spectroscopy RTP at the University of Warwick. We also thank Dr. Ben Breeze for assistance with Raman and UV–VIS measurements. R.P. acknowledges support from the European Research Council (grant 101163117, FemtoCharge).